\documentclass[12pt,a4paper]{article}
\usepackage{amssymb}
\usepackage{amsmath}
\usepackage[symbol]{footmisc}
\usepackage{amsthm}
\usepackage{amscd}
\usepackage{multirow}
\usepackage{amstext}
\usepackage{booktabs}
\usepackage{vmargin}
\usepackage{float}
\usepackage{fancyhdr}
\setpapersize{A4}
\setmarginsrb{3.0cm}{1.5cm}{3.0cm}{3cm}{1cm}{.5cm}{2cm}{2cm}

\usepackage{amsfonts}
\usepackage[latin1]{inputenc}
\usepackage[T1]{fontenc}
\usepackage[american]{babel}
\usepackage{graphicx}
\usepackage{bbm}
\usepackage[mathscr]{euscript}
\usepackage{mathrsfs}
\usepackage{stmaryrd}
\usepackage{soul}
\usepackage{siunitx}
\usepackage{xspace}
\usepackage[draft,danish]{fixme}
\usepackage{mathtools}
\usepackage{dsfont}
\usepackage{url}
\usepackage[ps,all,dvips]{xy}
\usepackage[shortlabels,inline]{enumitem}
\setlist[enumerate]{font=\textnormal}

\SelectTips{cm}{12}
\CompileMatrices

  \makeatletter
 \useshorthands{"}
 \defineshorthand{"-}{\nobreak-\bbl@allowhyphens}
 \makeatother

\theoremstyle{plain}

\theoremstyle{definition}

\DeclareMathOperator*{\argmax}{arg\,max}

\theoremstyle{definition}

\theoremstyle{remark}

\newcommand{\distEq}[1][]{%
	\overset{%
		\smash{%
			\raisebox{-0.5pt}{%
				\ $
				\scriptscriptstyle%
				d%
				$
			}%
		}%
	}{%
		=%
	}
}

\newcommand{\devnull}[1]{}

\numberwithin{equation}{section}

\title{A surrogate model for estimating extreme tower loads on wind turbines based on random forest proximities}
\author{Mikkel Slot Nielsen\footnote{Department of Statistics, Columbia University, USA. E-mail: m.nielsen@columbia.edu.}
	 \and Victor Rohde\footnote{Department of Mathematics, Aarhus University, Denmark.} 
}
\date{\ }
	
\begin{document}
\maketitle

\begin{abstract}
In the present paper we present a surrogate model, which can be used to estimate extreme tower loads on a wind turbine from a number of signals and a suitable simulation tool. Due to the requirements of the International Electrotechnical Commission (IEC) Standard 61400-1, assessing extreme tower loads on wind turbines constitutes a key component of the design phase. The proposed model imputes tower loads by matching observed signals with simulated quantities using proximities induced by random forests. In this way the algorithm's adaptability to high-dimensional and sparse settings is exploited without using regression-based surrogate loads (which may display misleading probabilistic characteristics). Finally, the model is applied to estimate tower loads on an operating wind turbine from data on its operational statistics. 
\end{abstract}

\section{Introduction}\label{introduction}

The aim of this paper is to construct a surrogate model for tower loads on wind turbines based on simulation. By using operational statistics (signals) from a turbine the surrogate model should be able to estimate load distributions and, in particular, extreme loads such as the 50-year return load. Surrogate models are widely employed in engineering science and can, among other things, be used to avoid computational costs caused by additional simulations and to impute missing outputs in measurements. The applications of surrogate models in the context of wind turbines are many, e.g., they are used to characterize energy production and lifetime equivalent fatigue loads (\cite{murcia2018uncertainty}), design and optimize turbine blades (\cite{meng2019structural}), and design the rotor (\cite{sessarego2016aerodynamic}). Estimation of extreme loads is also a well-studied area, particularly due to the IEC standards (see, e.g., \cite{ragan2008statistical,veers2001extreme}). However, the possibility of building extreme load estimates on top of surrogate loads has, to the best of our knowledge, not been investigated.

The model must be able to (i) impute surrogate loads which approximately have the true distribution conditional on the signals, and (ii) handle a high-dimensional feature space (many signals). Since it is important to extract other distributional properties than simply the mean of the conditional load, standard prediction-based imputation will not work (\cite{scheffer2002dealing}). Under the assumption of strong ignorability (in particular, the load distribution is the same for both measurements and simulations conditional on the signals), matching observations with simulations based on signals is a reasonable way to meet (i). However, due to the curse of dimensionality, classical matching techniques, such as $k$-nearest neighbors, will only work well if the dimension of the feature space is small (\cite{beyer1999nearest}).

In this paper we propose imputing surrogate loads using random forest proximities. Based on data from an operating wind turbine with several operational statistics and a simulation tool we demonstrate that, by matching signals with simulated values, the random forest proximities can successfully impute tower loads. Moreover, these imputations are precise enough to allow for subsequent estimation of extreme events. Random forest algorithms have become an extremely popular tool in the analysis of high-dimensional data and, by some authors, viewed as the most successful general-purpose algorithm in modern times (\cite{howard2012two}). A random forest is an ensemble of decision trees; given a tree-building mechanism, a large number of decision trees is constructed and, finally, trees are aggregated to form the forest. Arguably, the most popular forest is the one introduced by Breiman \cite{breiman2001random} where each tree is constructed in the following way:

\begin{enumerate}[(i)]
	\item Draw a bootstrap sample (with replacement) from the original data.
	
	\item Form a finer and finer partition of the feature space by recursively performing splits perpendicular to the axes. Each split is performed by maximizing impurity decrease over a randomly chosen subset of signals of fixed cardinality (see Section~\ref{model} and \cite{breiman1984} for details).
	
	\item The partitioning procedure continues as long as each leaf contains a prespecified, and often small, number of observations.
	
\end{enumerate}
\noindent The reason that this algorithm succeeds in large dimensions is that splits are placed in a greedy manner: the split is placed to decrease the impurity within each node as much as possible. As a consequence of this, most splits will be placed along important signals, disregarding directions in the feature space which do not contribute to explaining (the mean of) the output. While the forest is often used for prediction by either voting or averaging across trees, it also induces its own measure of proximity: given two points in the feature space, the proximity is the proportion of trees, where they fall into the same leaf (\cite{biauRF}). Using this measure it is possible to perform reasonable matching of signals even when the ambient dimension of the space is large. Clustering and matching of signals using random forest proximities have been successfully applied in several studies (see, e.g., \cite{pierola2016ensemble,wang2013network,zhao2016propensity}).

The data set consists of $19\: 976$ measurements from a wind turbine under normal operation over a period from February 17 to September 30, 2017. Each measurement contains the state of 40 signals corresponding to four types of 10-minute operational statistics measured by 10 different sensors on the turbine. Specifically, the signals are the 10-minute maximum, minimum, mean and standard deviation of blade pitch angles (BLdA, BLdB and BLdC), electrical power (EP), generator speed (GenSp), pitch rates (PR1, PR2 and PR3), and tower top downwind and lateral acceleration (TT\_dw and TT\_lat). The interest will be in assessing the distribution of the 10-minute maximum downwind bending moment of the tower top, middle and base. Since this particular turbine is part of a measurement campaign, measurements of these loads are in fact available, and they will be used to assess the performance of the model. The model is based on $50\: 606$ simulated values of both signals and loads. 

Section~\ref{model} gives a mathematical introduction to random forests and their associated proximity measure. In relation to this, it is discussed how different components affect the proximity measure and, in particular, how to construct the forest so that its induced proximity suits the particular problem in question. Compared to the rest of the paper, this section is rather technical and, hence, the practitioner may turn directly to the application, which is treated in the subsequent sections. Section~\ref{simulationValidation} concerns some initial preparations of the data, such as detecting outliers and checking that simulated values of signals roughly overlap measured values, and Section~\ref{extremeEvents} presents the estimated load distributions and compares them to the true (empirical) distributions. Finally, Section~\ref{discussion} concludes the paper and points out interesting directions for future research.

\section{Random forests, the induced proximity measure and matching}\label{model}
In the following we introduce the random forests as well as their proximity measure and discuss their specifications. For details on their theoretical foundations beyond this, see e.g. \cite{biauRF,breiman2001random} or \cite[Chapter~15]{elementsLearning} and references therein. 

Let $X\in \mathcal{X}\subseteq \mathbb{R}^p$ be a vector of signals and $Y\in \mathbb{R}$ be the output variable satisfying $\mathbb{E}[Y^2]<\infty$. Suppose that we have $n$ observations $\mathcal{D}_n = \{(X_1,Y_1),\dots, (X_n,Y_n)\}$ available as training data. A (fully grown) \emph{decision tree} $r_n = r_n(\: \cdot \: ;\mathcal{D}_n)$ can be associated to a partition $(A_i^n)_{i=1}^n$ of $\mathcal{X}$ where each $A_i^n$ contains exactly one of the data points $\{X_1,\dots, X_n\}$. It is obtained recursively by starting from $A^1_1=\mathcal{X}$ and then, given that $A_i^k$ is a node containing at least two data points, a split direction $j \in \{1,\dots, p\}$ and position $\tau \in \{x^{(j)}\, :\, x \in A^k_i\}$ are chosen and $(A^{k+1}_\ell)_{\ell =1}^{k+1}$ is defined as $(A^k_\ell)_{\ell =1}^k$, but where $A_i^k$ is replaced by $A^k_i \cap \{x\, :\, x^{(j)} \leq \tau\}$ and $A^k_i \cap \{x\, :\, x^{(j)} > \tau\}$. Given a vector of signals $x$, the decision tree returns either the average (regression) or the majority vote (classification) of the observations which fall into the same node $A_n (x)$ in the partition as $x$. Throughout the paper, we will follow the lines of \cite{breiman2001random} for CART decision trees and choose the feasible split $(i^\star, \tau^\star)$ that maximizes the empirical impurity decrease, i.e., $(i^\star, \tau^\star)$ is the maximizer of
\begin{align}
\begin{aligned}\label{CARTsplit}
L_n (i,\tau) &= \sum_{j\colon X_j \in A} (Y_j - \bar{Y}_n(A))^2 - \sum_{j\colon X_j \in A\cap \{x\, :\, x^{(i)} \leq \tau\}} (Y_j - \bar{Y}_n (A\cap \{x\, :\, x^{(i)}\leq \tau\}))^2\\
&\quad - \sum_{j\colon X_j \in A\cap \{x\, :\, x^{(i)} > \tau\}} (Y_j - \bar{Y}_n (A\cap \{x\, :\, x^{(i)}> \tau\}))^2
\end{aligned}
\end{align}
over $i\in \mathcal{M}_A$ and $\tau \in \{x^{(i)}\, :\, x \in A\}$. Here $A\subseteq \mathcal{X}$ is a generic node, $\bar{Y}_n (A) = \frac{1}{\vert \{j\, :\, X_j \in A\}\vert}\sum_{j\colon X_j \in A} Y_j$ is the local average, and $\mathcal{M}_A$ is a random subset of $\{1,\dots, p\}$ of fixed cardinality $m\coloneqq\vert \mathcal{M}_A\vert$ defining the set of feasible split directions in $A$. In particular, the CART trees are built in a greedy manner: splits are placed along directions which contribute the most to explaining the variation of $Y$. The degree of greediness can be controlled by $m$.

A \emph{random forest} $\bar{r}_{n,M}$ (of size $M$) is simply a collection of decision trees,
\begin{equation*}
\bar{r}_{n,M}  = \{r_n^i \, :\, i=1,\dots, M\}.
\end{equation*}
All $M$ decision trees are assumed to be grown by the same set of rules, so that their diversity stems from an injected exogenous randomness only. More precisely, one assumes that $r^i_n = r_n (\: \cdot \: ; \mathcal{D}_n,\Theta_i)$ where $\Theta_1,\dots, \Theta_M$ are i.i.d.\ copies of some random variable $\Theta$. The variable $\Theta$ could include decisions concerning split direction, position and node, resampling schemes, choice of split among a set of candidates, etc. For Breiman's original algorithm outlined in Section~\ref{introduction}, $\Theta$ will include the bootstrap step as well as the random selection $\mathcal{M}_A$ of signals performed in each node prior to a split. The \emph{proximity} $\rho_{n,M} (x,\tilde{x})$ of two set of signals $x,\tilde{x}\in \mathcal{X}$ with respect to $\bar{r}_{n,M}$ is given by the frequency of trees in the forest where they share the same leaf:
\begin{equation}\label{proximity}
\rho_{n,M} (x,\tilde{x}) = \frac{1}{M}\sum_{i=1}^M \mathds{1}_{A_n(x;\mathcal{D}_n,\Theta_i)}(\tilde{x}).
\end{equation}
While \eqref{proximity} is the measure implemented in practice, the choice of $M$ is set by the modeler and only restricted by computational capabilities, so it is common practice to use the approximation $\rho_{n,M} (x,\tilde{x}) \approx \rho_n(x,\tilde{x})\coloneqq \mathbb{P}(\tilde{x}\in A_n (x;\Theta,\mathcal{D}_n)\mid \mathcal{D}_n)$ for the sake of interpretation. In words, $\rho_n (x,\tilde{x})$ is the probability that two points fall into the same leaf of a tree grown by using the randomization scheme $\Theta$ given the data $\mathcal{D}_n$. While the proximity measure is difficult to work with from a theoretical point of view for many types of forests (\cite{biauRF}), the intuition is clear:
\begin{enumerate}[(i)]
	\item\label{greedy} If trees are built in a greedy manner, most splits are placed along important signals, so $\rho_n (x,\tilde{x})$ can be large even though the two points $x$ and $\tilde{x}$ are far from each other in Euclidean distance as long as they mainly differ along redundant directions of the feature space.
	\item\label{nonGreedy} If trees are built on many randomly placed splits, $\rho_n (x,\tilde{x})$ is likely to be large only if $x$ and $\tilde{x}$ are close to each other along \emph{all} directions of the feature space.
\end{enumerate}
The proximity-based match $Y(x)$ associated to a vector $x$ of signals is the output $Y_i$ among the training data $\mathcal{D}_n$ where the input $X_i$ has the largest proximity when paired with $x$, i.e., $Y(x) = Y_{i^\star}$ where $i^\star = \argmax_i \rho_{n,M}(x,X_i)$. The forest should be built in such a way that the distribution of $Y(x)$ approximately coincides with conditional distribution of $Y$ given $X=x$. In the application below we will have $k$ measurements $\mathcal{D}^\prime_k = \{X^\prime_1,\dots, X^\prime_k\}$ of operational statistics operational from a wind turbine and $\mathcal{D}_n$ will be the simulations. The corresponding proximity based matches $Y^\prime_1,\dots, Y^\prime_k$ (the surrogate loads) are then used to estimate the CDF of the loads and, by extrapolation, extreme loads. For these results to be reliable it is important that the conditional distribution of $Y$ given $X=x$ is the same for both simulations and measurements.

We will follow the lines of Breiman's random forest and use (fully grown) CART trees in the application below, but to reduce the computational costs of implementing the algorithm we skip the bootstrap step. In particular, this means that $\Theta$ consists solely of the random selection of signals in each node and that the effects of \ref{greedy} and \ref{nonGreedy} must be balanced by the choice of $m$. On one hand, to be useful in a high-dimensional setting, $m$ should be large enough to ensure that important signals can be chosen most of the time. On the other hand, the CART splitting criterion \eqref{CARTsplit} is particularly suited for matching conditional means and, thus, it may treat signals which only impact higher order moments of $Y$ (e.g., its variance) as being redundant. Since we will be interested in obtaining an output $Y(x)$ which has the same distribution (and not only the same mean) of $Y$ given $X=x$, this suggests choosing $m$ small enough to ensure that, eventually, splits will be placed along any direction. While $m$ can be used as a tuning parameter by the modeler, we will use the default value $m= \lceil p/3 \rceil$ (cf.\ \cite{biauRF}) which seems to deliver good performance.

\section{Validating the simulation environment}\label{simulationValidation}
For the simulation environment to work it is necessary that the observed signals could as well have been simulated, i.e., the simulations overlap the measurements. To be a bit more precise, if $X \in \mathbb{R}^p$ is the vector of signals and $Z\in \{0,1\}$ describes whether we draw a measurement ($Z=1$) or not ($Z=0$), then we must have that 
\begin{equation}\label{overlap}
\mathbb{P}(Z=0\mid X)>0.
\end{equation}
This assumption is part of the strong ignorability assumption introduced by Rosenbaum and Rubin \cite{rosenbaum1983central} in relation to estimating heterogeneous treatment effects. 

\begin{figure}[htbp]
	\centering
	\includegraphics[width=0.8\textwidth,clip, trim={1cm 0.5cm 1.5cm 1cm}]{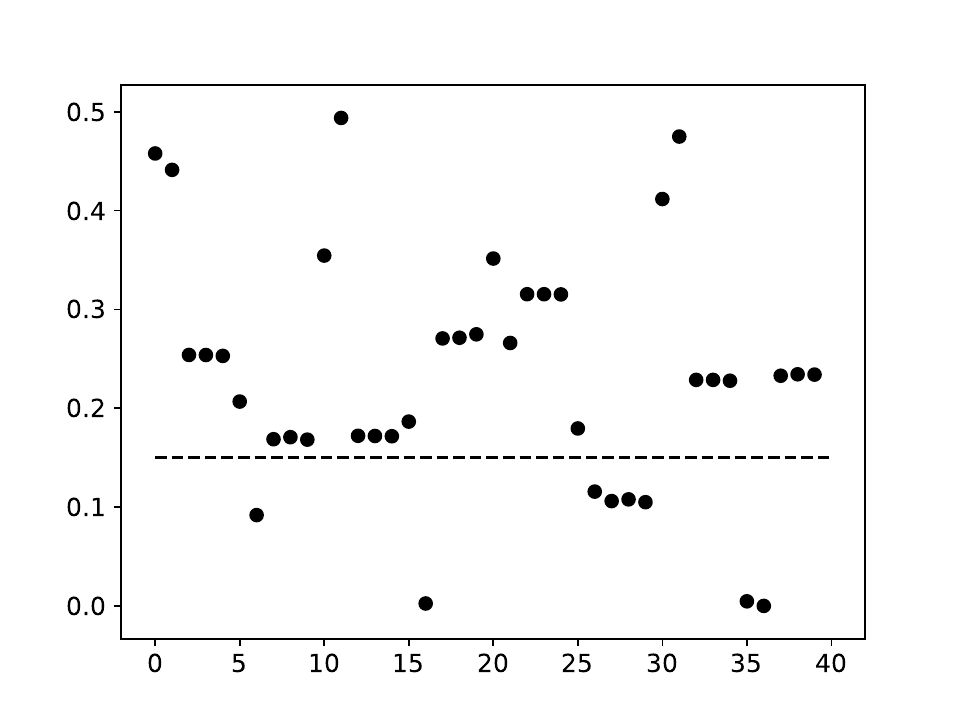}
	\caption{Averages of marginal propensity scores (second axis) over measurements for the 40 different signals (first axis). The threshold is 0.15.}\label{propScores}
\end{figure}

\begin{figure}[htbp]
	\centering
	\includegraphics[width=\textwidth,clip, trim={0cm 1.5cm 2.5cm 2cm}]{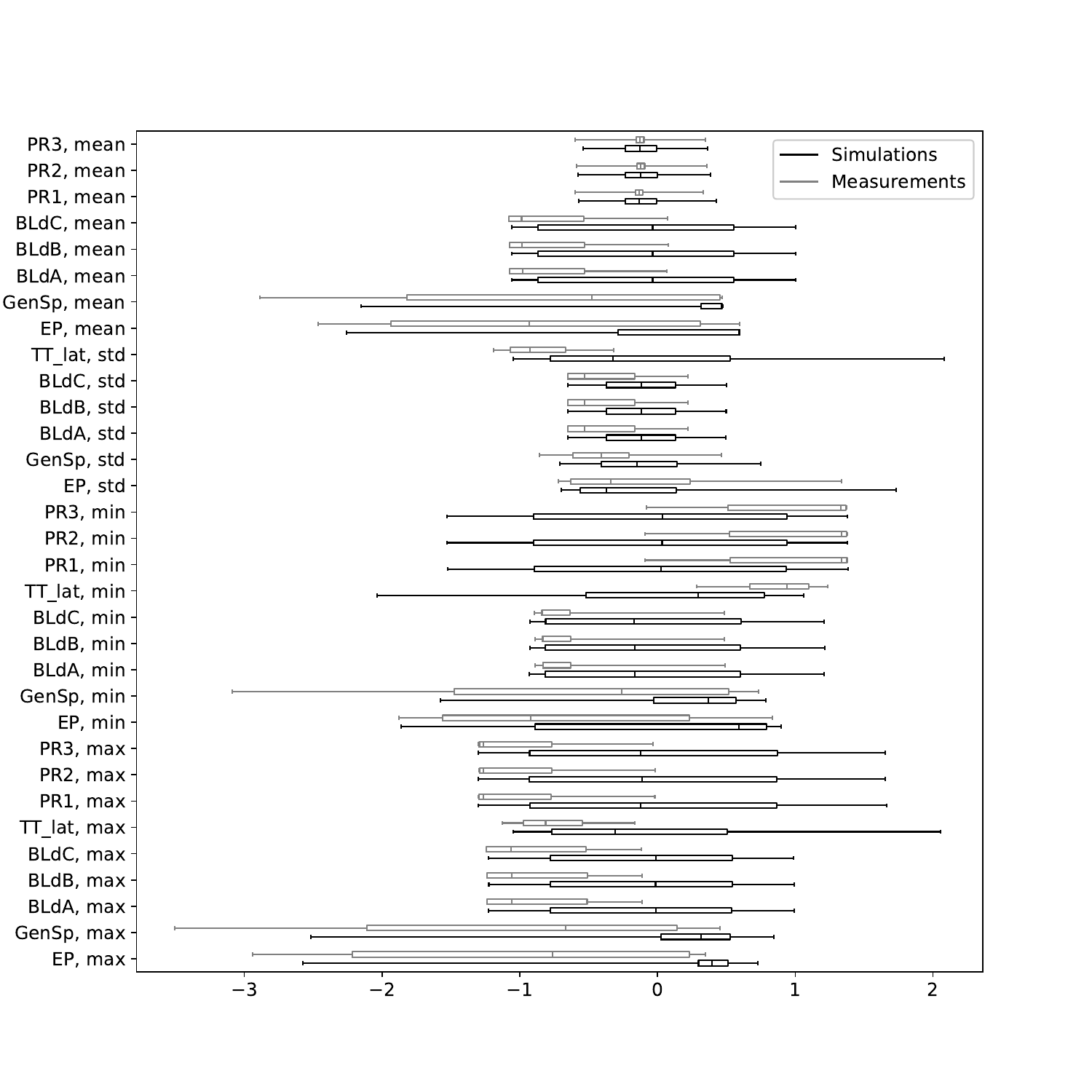}
	\caption{Box plots for the 32 included signals indicating the 0.05, 0.25, 0.50, 0.75 and 0.95 quantiles of their marginal distribution within simulations and measurements, respectively.}\label{propAbove}
\end{figure}

To accommodate \eqref{overlap} we initially detect and exclude signals which, to a great extent, can separate measurements from simulations. These are detected by first obtaining an estimate $\widehat{\rho}_i (x)$ of the marginal propensity score $\mathbb{P}(Z=0\mid X^{(i)}=x)$ by running a logistic regression and, next, computing the average $\bar{\rho}_i$ of $\widehat{\rho}_i (x)$ across all measured values $x$ of $X^{(i)}$. If $\bar{\rho}_i$ is close to 0, signal $i$ can almost separate the two environments and should be excluded from the analysis. In Figure~\ref{propScores} we plot $\bar{\rho}_i$ for the 40 signals. We see that the value is practically 0 for three signals (TT\_dw:mean, TT\_dw:min and TT\_lat:mean), and five more signals (PR1:std, PR2:std, PR3:std, TT\_dw:max and TT\_dw:std) have values around 0.1. It is the modeler's task to set an appropriate threshold; we choose it to be 0.15 and thereby exclude eight signals. To visualize the difference in the marginal distribution of simulations and measurements for the 32 included signals we provide box plots in Figure~\ref{propAbove} representing their 0.05, 0.25, 0.50, 0.75 and 0.95 quantiles. Although the masses are concentrated in different regions for some signals, it seems that, overall, simulations overlap measurements on the marginal~level. 

\begin{figure}[htbp]
	\centering
	\includegraphics[width=\textwidth, clip, trim={2.5cm 0cm 2.5cm 0cm}]{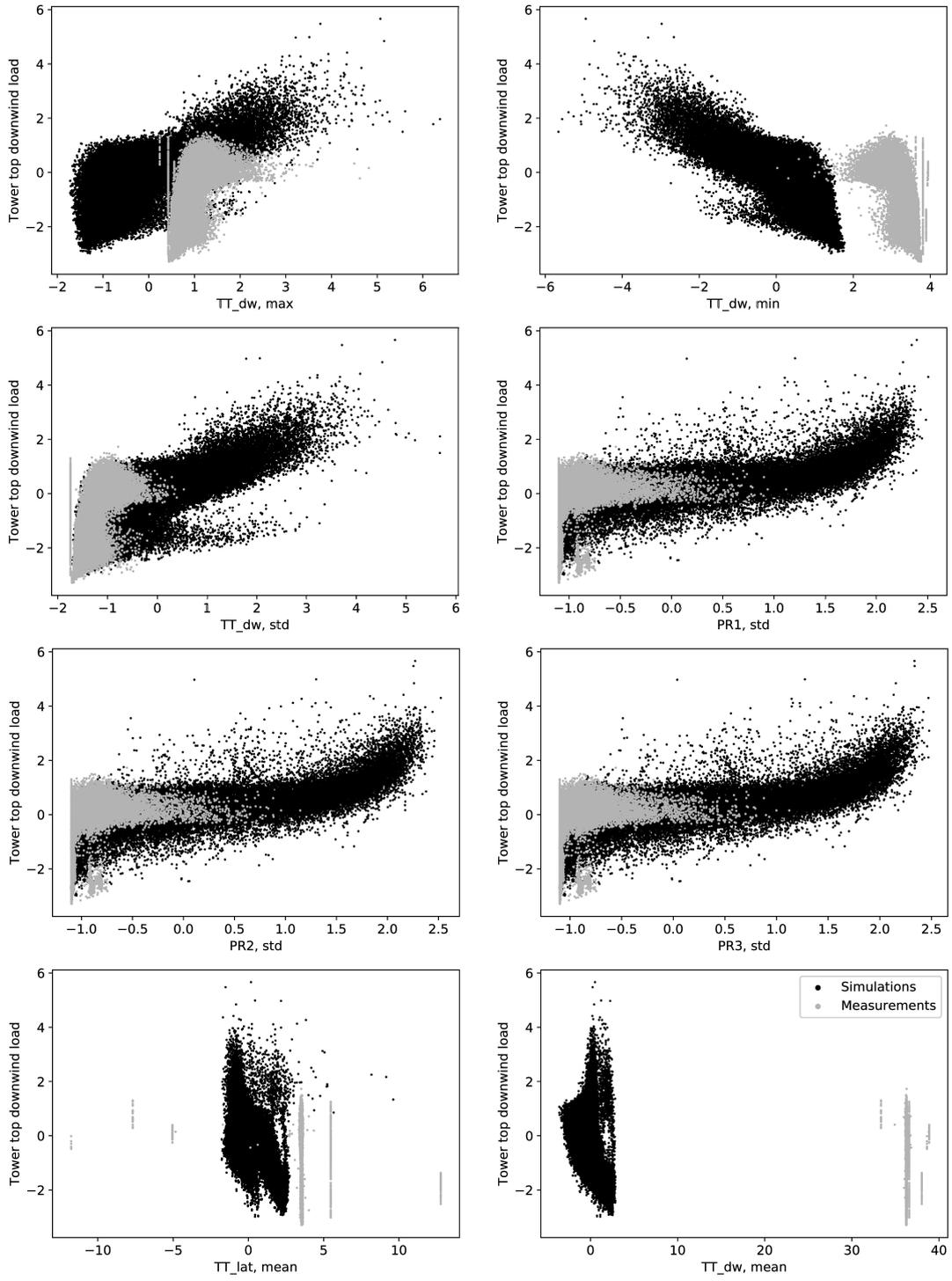}
	\caption{Plots of each of the excluded signals against tower top downwind load.}\label{input_selection}
\end{figure}

On the other hand, in Figure~\ref{input_selection} we have plotted the excluded signals against the tower top downwind load to illustrate the mismatch between measured and simulated values of those. Specifically, these plots show that the set of measurements is separated from the set of simulations for the three most critical signals, while it is near (and, in some regions, crossing) the boundary for the remaining five excluded signals. Another important observation is that many of the measurements of the excluded signals seem to be almost constant forming strange-looking vertical lines in the two-dimensional plots. Therefore, to a large extent, we suspect the mismatch to be caused by logging errors on these sensors, which supports the decision of discarding them from the subsequent analysis.

On top of this exclusion we identify and exclude measurements which are outliers relative to the simulations. Outlier detection methods have received a lot of attention over decades and, according to Hodge and Austin \cite{hodge2004survey}, they generally fall into one of three classes: unsupervised clustering (pinpoints most remote points to be considered as potential outliers), supervised classification (based on both normal and abnormal training data, an observation is classified either as an outlier or not) and semi-supervised detection (based on normal training data, a boundary defining the set of normal observations is formed). 
We will be using the so-called isolation forest for outlier detection, which belongs to the first class. An isolation forest consists of trees built by splitting along random directions and positions until the simulations are isolated, or until the tree has reached a certain depth. The degree of anomaly for a given measured signal $x$ with respect to a tree is inversely related to the number of edges that $x$ traverses before it reaches the leaf, and the isolation forest is an average across trees. The intuition behind the isolation forest is that outliers should be easy to isolate. The motivation for choosing this algorithm is, at least, three-fold: (i) it tends to work well in high-dimensional and sparse settings, (ii) it is computationally feasible with a linear time complexity, and (iii) it has shown favorable performance relative to other state-of-the-art methods (e.g., $k$-nearest neighbor based methods). For details about the isolation forest and general outlier detection methods, see \cite{liu2008isolation,liu2012isolation} and \cite{ben2005outlier,campos2016evaluation,hodge2004survey,zimek2012survey}, respectively.

\begin{figure}[htbp]
	\centering
	\includegraphics[width=\textwidth, clip, trim={5.5cm 1.5cm 4.7cm 3cm}]{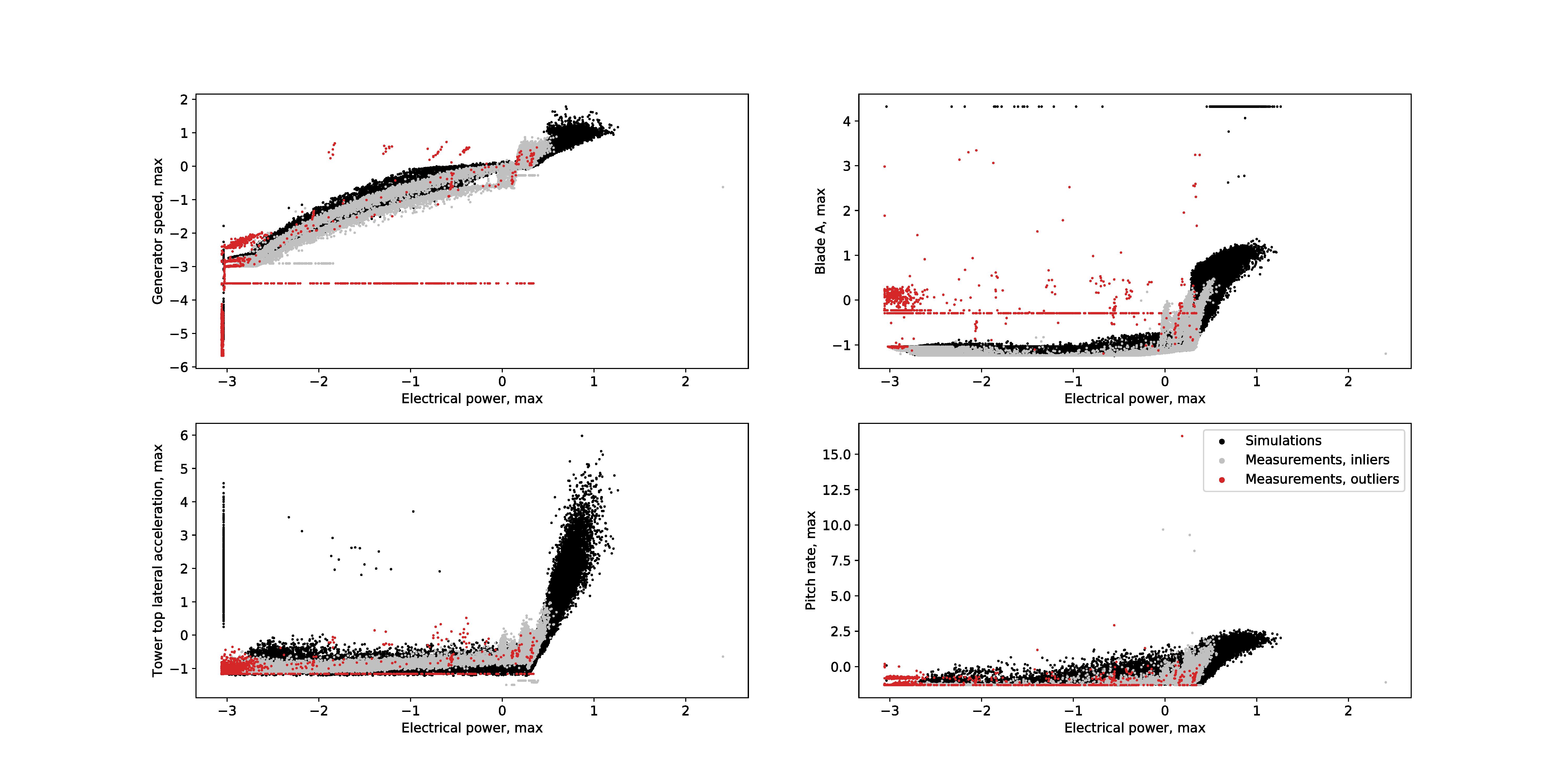}
	\caption{Two-dimensional projections of the signals. These plots correspond to EP:max against GenSp:max (upper left), BLdA:max (upper right), TT\_lat:max (lower left) and PR1:max (lower right).}\label{outliers}
\end{figure}

By evaluating the isolation forest on each measurement we obtain an anomaly ranking which is used to exclude a certain proportion $\alpha$ of the most suspicious measurements. We choose $\alpha = 0.05$ and check the performance of the algorithm through a visual inspection of the two-dimensional projections of the signals. In Figure~\ref{outliers} we have plotted four of these projections. Generally, the algorithm seems to identify most measurements which are not surrounded by simulations in the two-dimensional spaces, i.e., the majority of gray points are surrounded by black points. This statement can be further supported by plotting the simulations on top of measurements (and not only the other way around as in Figure~\ref{outliers}). In the following we will discard the outliers without addressing further how to handle these. However, one should observe that many of the observations, which are labeled as outliers, seem to form straight lines in Figure~\ref{outliers}. This indicates the presence of logging errors for one or more of the sensors and, thus, should be handled as a missing data problem.

\section{Estimating tower loads}\label{extremeEvents}
\begin{figure}[htbp]
	\centering
	\includegraphics[width=\textwidth, clip, trim={4.5cm 1.5cm 4.7cm 2.5cm}]{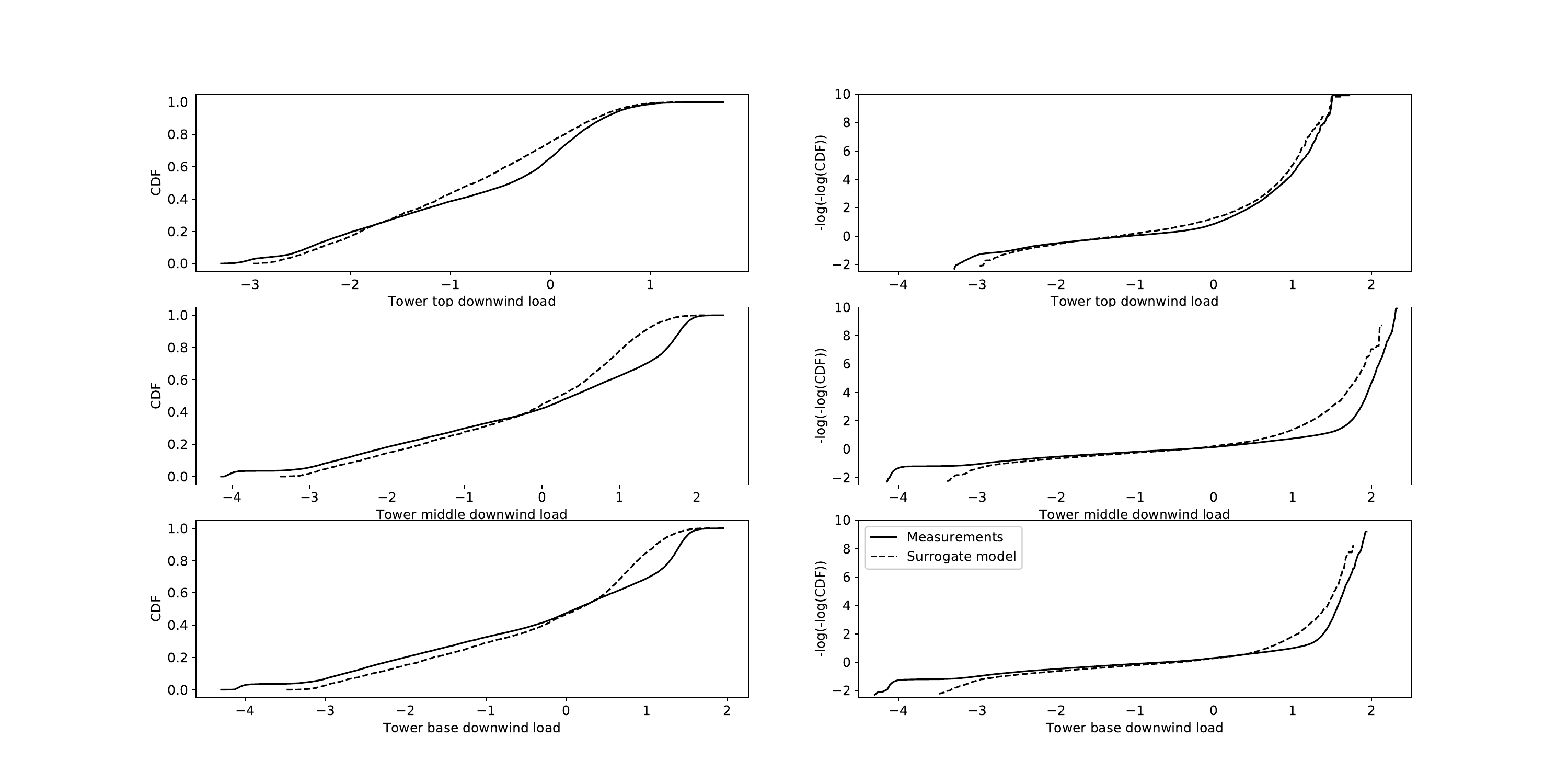}
	\caption{Empirical and surrogate CDFs (first column) and their corresponding $-\log(-\log(\cdot))$ transforms (second column) of tower top (first row), middle (second row) and base (third row) downwind load.}\label{loadCDF}
\end{figure}

After having removed the eight signals in Section~\ref{simulationValidation}, which separate measurements from simulations, and discarded the worst outliers in the measurements detected by the isolation forest, we can now impute tower loads by using the random forest proximity measure and the corresponding matching procedure explained in Section~\ref{model}. In particular, we use the proximity measure induced by Breiman's random forest built on 500 fully grown decision trees, and we skip the bootstrap step in the tree-growing procedure. Figure~\ref{loadCDF} shows the empirical CDF based on actual loads and the surrogate CDF based on imputed loads as well as the $-\log (-\log (\: \cdot \:))$ transform of the CDFs. This transform is natural to consider since loads reflect 10-minute maxima and, thus, the corresponding load distribution should, under suitable conditions, be close to a distribution of GEV type (cf.\ the Fisher--Tippett--Gnedenko theorem). Indeed, the right tail of the $-\log (-\log (\: \cdot \:))$ transform of a GEV distribution is either strictly convex (Weibull), strictly concave (Fréchet) or linear (Gumbel). Another reason for plotting this transform is that it puts most attention to the tail of the distribution, which is the important part when estimating extreme loads. To estimate rare (e.g., 50-year) return loads, one will need to extrapolate the tail of the surrogate CDF, e.g., by fitting the tail of a GEV distribution. While we have found promising results in this direction with errors on the 50-year return loads in the range \SIrange{5}{10}{\percent}, the performance relies heavily on the employed extrapolation method as well as how much of the tail that is used for extrapolation. Consequently, we have chosen not to include such an analysis here. For details on the extrapolation step, see \cite{ragan2008statistical}. Returning to the plots of Figure~\ref{loadCDF} it appears that the surrogate CDF for the tower top loads mainly differs from the empirical CDF on the interval between $-1$ and $0$, while for the middle and base load the critical region is between $1$ and $2$. Although this difference is also reflected in the $-\log (-\log (\: \cdot \:))$ transform, the plots indicate that extreme load estimation through extrapolation should generally work well, in particular for the tower top. It should be mentioned that plots of similar precision can be obtained for the lateral loads, except for the tower top which seems to be more challenging.

\section{Conclusion and future research}\label{discussion}
In this paper we demonstrated how random forest proximities can be used to obtain surrogate loads and discussed how they can lead to promising estimates of extreme loads. One of the key components of this algorithm is that trees are grown in a greedy manner, and hence it adapts to sparse settings where only few of the included signals have significant explanatory power on loads. Another important component is that loads are imputed by finding the best simulation in terms of proximity and not by prediction (regression). 

However, as pointed out in Section~\ref{extremeEvents}, there is indeed room for improvement, in particular for the tower middle and base downwind loads and for the tower top lateral loads. While it is difficult to quantify how much is caused by a mismatch between the conditional load distribution of measurements and simulations, it would in any case be natural to consider refinements of the employed algorithm. An interesting direction for future research would be to use an alternative splitting criterion than the CART specification \eqref{CARTsplit} in order to favor signals that explain higher order moments of the conditional load distribution. The CART methodology is particularly tailored to detect signals which explain the conditional mean load. Related to this, one could consider other randomization schemes (other specifications of $\Theta$) and actively use the degree of randomness as a tuning parameter to control the diversity between trees and ensure splits on important signals which are not detected by the splitting criterion.

\section*{Acknowledgements}

We thank James Alexander Nichols from Vestas (Loads \& Control) and Jan Pedersen for fruitful discussions.

\section*{Funding}

This work was supported by the Danish Council for Independent Research under Grants 4002 - 00003 and 9056 - 00011B.

\bibliographystyle{plain}
\bibliography{bibliography}
\end{document}